\documentclass[twocolumn,showpacs,preprintnumbers,amsmath,amssymb]{revtex4}

\usepackage{graphicx}
\usepackage{dcolumn}
\usepackage{bm}

\begin{document}

\title{\large A study of the ferromagnetic transition
of $SrRuO_3$ in nanometer thick bilayers with $YBa_2Cu_3O_y$,
$La_{1.88}Sr_{0.12}CuO_{4-y}$, Au and Cr:\\
Signature of injected carriers in the pseudogap regime }

\author{G. Aharonovich, G. Koren and E. Polturak }
\affiliation{Physics Department, Technion - Israel Institute of
Technology Haifa, 32000, ISRAEL}

\email{gkoren@physics.technion.ac.il}
\homepage{http://physics.technion.ac.il/~gkoren}

\date{\today}
\def\bfig {\begin{figure}[tbhp] \centering}
\def\efig {\end{figure}}

\normalsize \baselineskip=4mm  \vspace{15mm}

\begin{abstract}
The hypothesis regarding the existence of uncorrelated pre-formed
pairs in the pseudogap regime of superconducting $YBa_2Cu_3O_y$ is
tested experimentally using bilayers of $YBa_2Cu_3O_y$ and the
itinerant ferromagnet $SrRuO_3$. In our study, we monitor the
influence of $YBa_2Cu_3O_y$ on $T_p$, the ferromagnetic ordering
temperature of $SrRuO_3$. Here, $T_p$ is the temperature of
maximum dM/dT or dR/dT where M and R are the magnetization and
resistance of $SrRuO_3$, respectively. We compare the results with
similar measurements carried out on bilayers of
$La_{1.88}Sr_{0.12}CuO_{4-y}$, $Au$ and $Cr$ with $SrRuO_3$. We
find that in bilayers made of underdoped 10 nm $YBa_2Cu_3O_y$/5 nm
$SrRuO_3$, the $T_p$ values are shifted to lower temperatures by
up to 6-8 K as compared to $T_p\approx 140$ K of the 5 nm thick
reference $SrRuO_3$ film. In contrast, in the other type of
bilayers, which are not in the pseudogap regime near $T_p$, only a
smaller shift of up to $\pm$2 K is observed. These differences are
discussed in terms of a proximity effect, where carriers from the
$YBa_2Cu_3O_y$ layer are injected into the $SrRuO_3$ layer and
vice versa. We suggest that correlated electrons in the pseudogap
regime of $YBa_2Cu_3O_y$ are responsible for the observed large
$T_p$ shifts.
\end{abstract}

\pacs{74.45.+c, 75.70.-i, 74.50.+r, 74.78.Bz}

\maketitle

There exists a large body of experimental evidence to date for the
existence of a pseudogap regime above $T_c$ in the high
temperature superconducting cuprates \cite{Timusk-Statt,Norman}.
Among the several possible models put forward to explain the
origin of this phenomenon, the precursor pairs scenario carries a
certain appeal. This model assumes the existence of pre-formed
pairs in the pseudogap regime which do not show phase coherence.
On lowering the temperature, these precursor pairs reach phase
coherence at the superconducting transition temperature $T_c$
\cite{Emery-Kivelson}. The pre-formed pair scenario is consistent
with many experimental results in the pseudogap regime, but
experimental verification for the existence of the elusive
pre-formed pairs is still lacking. In the present study we
investigate the proximity effect in the pseudogap regime using
bilayers of $YBa_2Cu_3O_y$ (YBCO) and the itinerant ferromagnet
$SrRuO_3$ (SRO). We observe that the proximity effect leads to a
large decrease of the ferromagnetic transition temperature.
Ferromagnetic order is incompatible with singlet Cooper pairs.
Consequently, the presumed pre-formed pairs are likely candidates
to explain this effect. Previous results on the proximity effect
using superconducting-ferromagnetic (SF) bilayers and multilayers
have been reported mostly for YBCO and a manganite such as
$La_{2/3}Ca_{1/3}MnO_{3}$ (LCMO) showing giant magnetoresistance.
It was shown by magnetization measurements, that the temperatures
where the magnetic moment saturates below the superconducting
$T_c$, and of the onset of the ferromagnetic transition at
$T_{Curie}$ above $T_c$, are both suppressed with increasing
thickness of the superconducting layer \cite{Soltan}. The results
above $T_c$ were interpreted as due to a possible charge transfer
from the LCMO to the YBCO layer. Similar results on the
suppression of $T_{Curie}$ versus the YBCO thickness in
multilayers of YBCO/LCMO above $T_c$, were obtained also in
resistivity and susceptibility measurements \cite{Pena,Lopez}. The
opposite effect, where the superconducting $T_c$ is depressed by
the ferromagnetic layer is also possible due to pair breaking by
spin polarized carriers penetrating the superconductor
\cite{Soltan,Pena,Moraru}. No reference to the pseudogap or
preformed pairs role in this context was mentioned or discussed in
these studies.\\

In our experiment, epitaxial thin films and bilayers of SRO with
either YBCO, $La_{1.88}Sr_{0.12}CuO_{4-y}$ (LSCO), Cr or Au were
deposited \textit{in-situ} by laser ablation deposition on (100)
$SrTiO_3$ (STO) wafers of $10\times 10$ mm$^2$ area. For
reference, nominally identical single layer films of these
materials on STO were also prepared. All SRO, YBCO and LSCO layers
in the different heterostructures were oriented with their
\textit{c-axis} normal to the wafer. The SRO and LSCO layers were
prepared under the same deposition conditions as for obtaining
high quality YBCO films (100 mT of oxygen flow and at
780$^\circ$C). The Cr layer was deposited under vacuum at
30$^\circ$C, while the Au layer was deposited in 400 mT oxygen at
150$^\circ$C. Transmission electron microscope (TEM) images of
similar SRO films deposited on (100) STO show an atomically smooth
interface with the STO wafer and were found to grow in the
layer-by-layer mode up to at least 10 nm thickness
\cite{Char-TEM}. Fig. 1 shows an image of the surface morphology
of one of our SRO films measured by a scanning tunneling
microscope (STM). One can see that this film consists of a stack
of parallel and flat plates which are $\sim 60-90$ nm wide and
atomically smooth. These plates are separated by steps of one unit
cell height (the \textit{c-axis} of SRO), and are formed due to
the $\sim 0.5^\circ$ miscut angle of the STO substrate. Thus the
area onto which the cover layers in the bilayers with SRO are
deposited on is atomically smooth and almost flat. In view of the
small fraction of the total area of the film that the atomic steps
occupy, we can safely assume that the dominant contribution to the
properties of the bilayers originates in the flat areas. TEM
images of the YBCO/SRO interface are also atomically smooth
\cite{Habermeier-Physica-C}, and similar to the YBCO/LCMO
interfaces in super-lattices \cite{Habermeier-PRB}. We can
therefore conclude that the sharp and flat interfaces exclude any
inter-diffusion and chemical reactions which could affect our
results. Recently, epitaxial $SrTiO_3/LaAlO_3$ bilayers with
different termination layers at the interface were investigated
\cite{Nakagawa}. The results show that the effects of ionic and
charge compensation for the different cases play a major role in
enabling the creation of a sharp interface on a scale of a single
atomic layer. In that study however \cite{Nakagawa}, both
components of the bilayer are ionic and insulating, while in our
case, all the layers are metallic and conducting. Therefore,
charge compensation is taken care of automatically, and the
resulting interfaces are smooth and sharp. We also note that even
if only one layer in the bilayer is conducting like in the SRO/STO
case, the interface is
clearly sharp \cite{Char-TEM}.\\

\begin{figure}
\includegraphics[height=7cm,width=8cm]{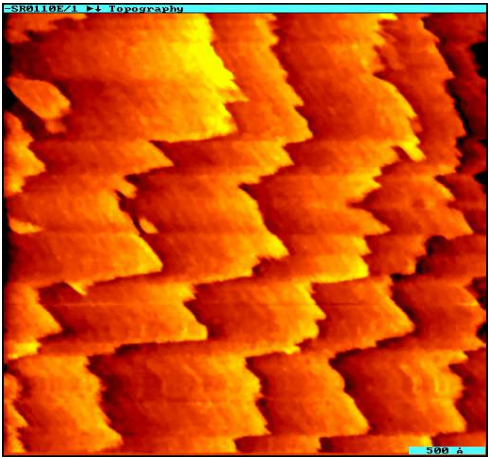}
\caption{\label{fig:epsart}(Color online) A scanning tunneling
microscope image of $300\times 300$ nm$^2$ area, showing the
surface morphology of a typical SRO film deposited on (100) STO
wafer. The observed plates are atomically smooth, and the steps
between them are of one unit cell height in the c-direction of SRO
(0.785 nm).}
\end{figure}

In order to minimize effects of wafer to wafer  variations between
different deposition runs, some bilayers were prepared
\textit{in-situ} together with the reference films in the
following way: first, the SRO layer was deposited on the whole
area of the wafer. Then a shadow mask made of an MgO wafer was
used to cover half of the sample area while the YBCO layer was
deposited on the other half. This resulted in an SRO reference
film on half the wafer and a bilayer on the other half. The
various single layer films and bilayers deposited on the whole
area of the wafer were not patterned. Wafers which were half
coated with a bilayer and half with the SRO film were patterned by
either wet etching or a scratch of a narrow stripe to separate the
bilayer from the reference film. A very convenient feature of our
experiment was that YBCO/SRO bilayers could be reannealed in $O_2$
to produce YBCO with different values of $T_c$ without affecting
the properties of the SRO. We note that the oxygen annealing
process was done at 450$^\circ$C which is about half the
deposition temperature, and was fully reversible. Namely, we could
switch back and forth between the different $T_c$ values of the
bilayers without any change in their properties. This again is
consistent with the absence of interdiffusion which would have
affected the properties of the bilayer in a progressive and
irreproducible manner. This convenient feature allowed us to
compare the proximity effects at different doping levels on the
same sample. Transport measurements were done by the standard
4-probe technique using gold coated contact tips. All in all,
about 30 bilayers and reference films were prepared and measured,
to establish the reliability and
reproducibility of our results.\\

In the present study, resistance versus temperature measurements
were used rather than direct magnetization measurements, because
of the higher sensitivity that can be obtained in measuring a few
nanometer thick SRO film (typically 5 nm). Theoretically, the
issue of resistive anomalies and peaks associated with
ferromagnetic transitions was discussed quite long ago
\cite{Langer,Helman}. Here we wish to show first, experimentally,
that near the ferromagnetic transition, the functional dependence
of dR/dT is very similar to that of dM/dT where R is the
resistance, T is the temperature and M is the magnetization of the
sample. For this, magnetization measurements as a function of
temperature were performed using a SQUID magnetometer, and
compared with measurements of R and dR/dT of the same sample. Fig.
2 shows the results of these measurements on a 200 nm thick SRO
film. One can see that on cooling down, the magnetization curve
starts rising at $T_{Curie}\approx 150\,K$ where the ferromagnetic
order sets in, and that this temperature coincides with the
corresponding sharp increase of both dR/dT and dM/dT. The
inflection point of M versus T yields a peak in dM/dT which also
coincides with the peak of dR/dT. We denote this temperature by
$T_p$. Thus a basic correspondence between dR/dT and dM/dT near
$T_{Curie}$ is well established. For the thin films of a few nm
thickness used in this study, the magnetization signal was too
small to measure. We therefore used measurements of dR/dT to
detect $T_p$, the midpoint of the ferromagnetic transition. \\

\begin{figure}
\includegraphics[height=6cm,width=8cm]{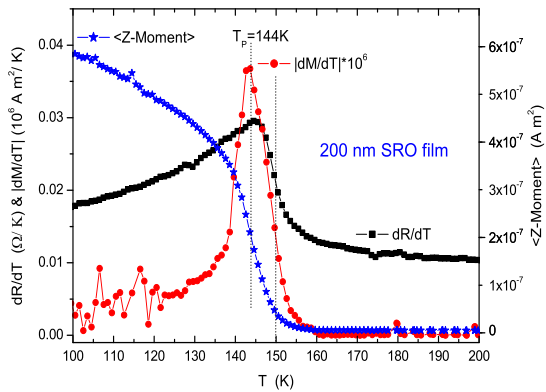}
\caption{\label{fig:epsart}(Color online) Temperature derivative
of the resistance versus temperature of a 200 nm thick SRO film,
together with the magnetization of the same film and its
temperature derivative. }
\end{figure}

The choice of thickness of the films in this study was done by
searching for the combination that would produce a large
observable effect on the SRO transition. In general, to achieve
good sensitivity, the resistances of the two components of the
bilayer should be comparable. Second, the thickness should be in
the range of the relevant penetration depths of the proximity
effect. If one of the layers is much thicker than the penetration
depth, the observable effect will be small. For example, with the
YBCO film much thicker than the SRO, the SRO transition was
suppressed to the point where an unambiguous identification of
$T_{p}$ became impossible. Finally, if the layers are too thin,
the number of carriers available for injection into the other film
is too small to produce an observable effect. For example, as
shown below, in bilayers of 7.5 nm YBCO on 5 nm SRO, any
suppression of the SRO transition temperature was smaller than the
variation of $T_{p}$ between different SRO films. In contrast, a
clear suppression was observed in bilayers of 10 nm YBCO on 5 nm
SRO. The thickness of the films used in this study indicates that
the range of the proximity effect is on the order of a few nm. In
view of the effects discussed above concerning the layers
thickness in the bilayers, we also conclude that the measured
results in the bilayers reflect mostly their bulk properties and
not the interface.\\

\begin{figure}
\includegraphics[height=6cm,width=8cm]{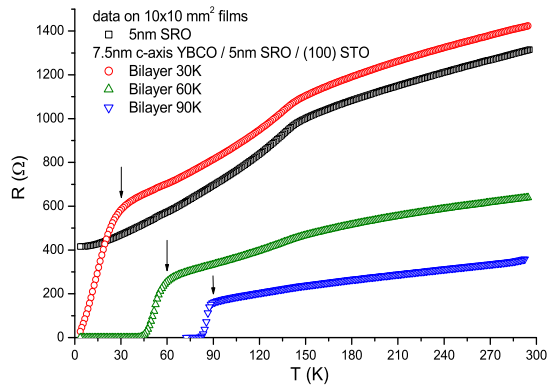}
\caption{\label{fig:epsart}(Color online) Resistance versus
temperature of a 5 nm SRO film, and of a 7.5 nm YBCO/5 nm SRO
bilayer reannealed under different conditions to produce YBCO with
$T_c$ of either 90, 60 or 30 K (see the arrows). }
\end{figure}

\begin{figure}
\includegraphics[height=6cm,width=8cm]{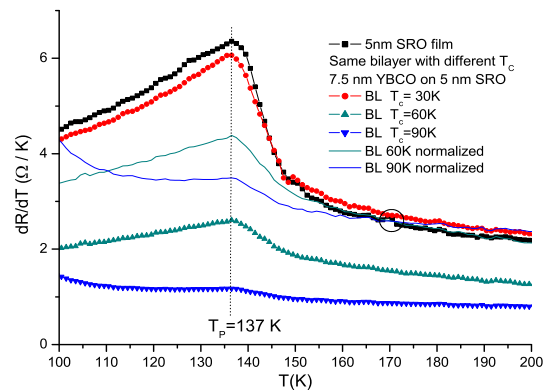}
\caption{\label{fig:epsart}(Color online) Temperature derivatives
dR/dT of the resistance versus temperature data of Fig. 3. The
curves without symbols represent the 60 and 90 K bilayers data
normalized to the reference SRO data in the temperature range
indicated by the circle.}
\end{figure}

Figs. 3 and 4 show the resistance versus temperature and the
corresponding temperature derivatives dR/dT of a 5 nm thick SRO
film, and of a 7.5 nm YBCO/5 nm SRO bilayer under three different
oxygenation levels. The value of $T_{p}$ in 5 nm thick SRO films
is typically around 137-140 K, slightly lower than in the bulk
(144 K, as seen in Fig. 2). It was established independently that
the SRO films are insensitive to the oxygen annealing conditions
used in the present study. In contrast, the YBCO layer in the
bilayer is very sensitive to the oxygen annealing conditions, and
in Figs. 3 and 4 we show the results for the 30, 60 and 90 K
phases of YBCO obtained on the same bilayer through repeated
annealing. We note that the oxygen annealing process is
reversible, as we can switch between different $T_c$ values with
reproducible transport results. This multiple annealing without
deterioration in the transport properties is apparently due to the
fact that the oxygen annealing temperature (450$^\circ$C) is much
lower than the deposition temperature (780$^\circ$C) where the
layered structure is formed. Fig. 4 shows that at temperatures
above 100 K, the 30 K bilayer data is almost coincident with that
of the SRO film. This is due to the high normal resistivity of the
30 K YBCO phase and indicates very little interaction between the
layers in this case. For the 60 and 90 K phases of YBCO this is no
longer the case. A suppression of the magnitude of $dR/dT$ near
$T_{p}$ is clearly observed. To account for the different
resistances of the differently oxygenated bilayers as in Fig. 3,
and for the sample to sample variability due to morphology
differences and so on, we generally normalized the dR/dT data to
that of the SRO reference film in the temperature range of 160-180
K above the ferromagnetic transition at $T_{Curie}$=150 K. In
cases where the dR/dT curves did not fully overlap in this regime,
normalization was done around 170 K, like in Fig. 4. The
normalized curves show that $T_p$, the midpoint of the
ferromagnetic transition (shown by the curves without symbols in
Fig. 4) remains constant at 137 K, but the overall magnitude of
the dR/dT signal is still significantly
suppressed in these bilayers. \\

\begin{figure}
\includegraphics[height=6cm,width=8cm]{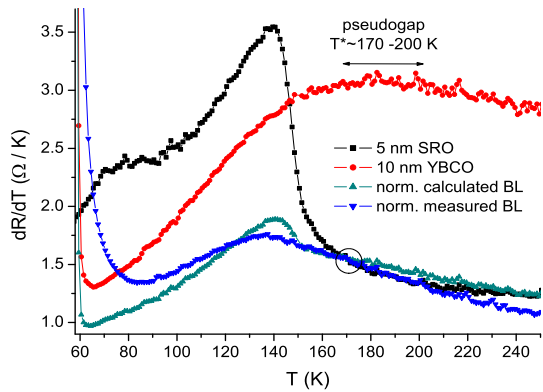}
\caption{\label{fig:epsart}(Color online) Temperature derivatives
dR/dT of the resistance versus temperature of a 5 nm thick SRO
film and a 10 nm thick 60 K YBCO film together with  the
normalized data of the measured and calculated results of 10 nm
YBCO/5 nm SRO bilayers. The bilayer curves are normalized to the
reference SRO data at 170 K as indicated by the circle.}
\end{figure}

In the next step, we prepared bilayers with a slightly thicker
YBCO layer. Fig. 5 shows the dR/dT data versus temperature of a
bilayer of 10 nm thick YBCO film on top of a 5 nm thick SRO layer,
together with the data of the corresponding reference films. Also
shown in Fig. 5 is the expected resistance of the same bilayer
structure, calculated when the two layers are assumed to be
noninteracting and behave like two independent resistors connected
in parallel (termed "calculated" in Fig. 5). The bilayer and films
were annealed in a low oxygen pressure (of 10 mTorr $O_2$ flow) to
produce the $T_c$=60 K phase of YBCO. At this doping level, the
pseudogap regime of YBCO sets in below $T^*$ of about 170-200 K
(this is seen as the broad shallow peak of the YBCO film in Fig.
5). As before, the SRO films are found to be insensitive to the
oxygen annealing pressure, with $T_{Curie}$ of 150 K. This is
easily seen by the change of slope at 150 K of the resistance
versus temperature curve of the SRO film in Fig. 3, and by the
corresponding sharp increase of dR/dT in Fig. 5 at 150 K. These
observations are in agreement with the results found in the
literature \cite{Char-SRO,Aronov-CARE}. Just below 150 K, the
measured dR/dT of the bilayer in Fig. 5 is clearly lower than the
calculated resistance of the corresponding noninteracting bilayer.
Thus one can conclude that the different layers do affect one
another in such a way that decreases their dR/dT below the
ferromagnetic transition temperature. Here, a shift of the
ferromagnetic peak temperature $T_p$ to lower temperatures is
observed, and this will be discussed in more detail next.\\

\begin{figure}
\includegraphics[height=6cm,width=8cm]{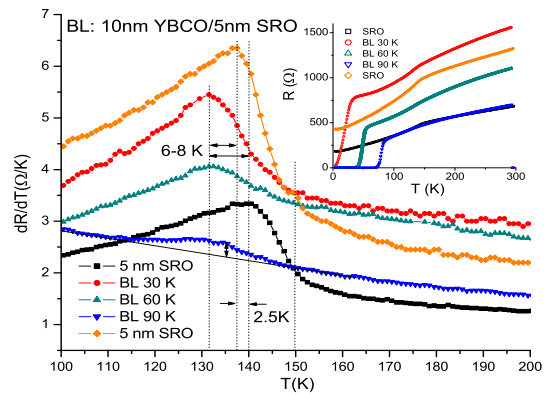}
\caption{\label{fig:epsart}(Color online) dR/dT versus temperature
of the same 10 nm YBCO/5 nm SRO bilayer reannealed to produce the
30, 60 and 90 K YBCO phases, together with the dR/dT data of two 5
nm thick SRO films of high and low resistances. The inset shows
the corresponding raw data of R versus T.}
\end{figure}

\begin{figure}
\includegraphics[height=6cm,width=8cm]{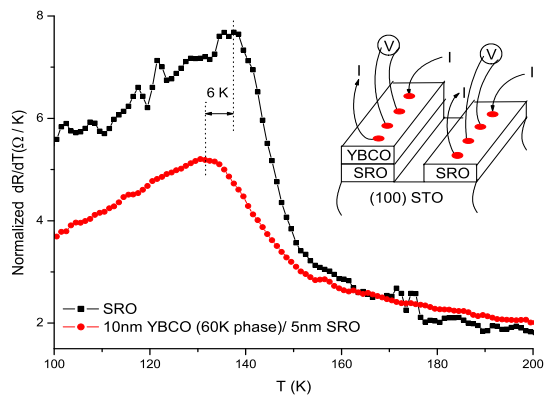}
\caption{\label{fig:epsart}(Color online) dR/dT versus temperature
of a 10 nm YBCO (60 K phase)/5 nm SRO bilayer normalized to its 5
nm thick SRO reference film on the same wafer. Normalization is
done in the 160-180 K temperature range. The inset shows a
schematic drawing of the bilayer and reference film on the wafer
together with the two four probe contacts configuration.}
\end{figure}

Figure 6 shows dR/dT results versus temperature of a 10 nm YBCO/5
nm SRO bilayer obtained on the same wafer by re-annealing in
oxygen to produce the 30, 60 and 90 K YBCO phases. A clear shift
to lower temperatures of 6-8 K is now observed in the peak
position of the ferromagnetic ordering temperature $T_p$, in the
bilayers of the 30 and 60 K YBCO phases, relative to the $T_p$
values of two reference SRO films with high and low resistances.
We verified by atomic force microscopy, that the different
resistances of the two 5 nm thick SRO films are related to
morphology changes in the films. This can be due to a slight
difference in the miscut angle of the (100) $SrTiO_3$ wafers or to
small film thickness variations in different deposition runs. To
minimize this large variability in the properties of the SRO
films, we prepared additional samples with the reference SRO film
on half the area of the wafer and the bilayer on the other half as
described in the experimental part before. This geometry is
depicted in the inset of Fig. 7. The results of the normalized
dR/dT in such a sample are shown in Fig. 7. Basically, the results
of Fig. 6 are now reproduced in Fig. 7 in a bilayer with the 60 K
YBCO phase where again, a large 6 K shift of $T_p$ is found. This
time however, the results were obtained with two SRO films (one in
the bilayer and the other of the reference film) which were
prepared in the same deposition run and on the same wafer.
Therefore, the variability in the SRO properties in this case is
minimal, and the reliability and reproducibility of the $T_p$
shift is well established. A smaller shift of $T_p$ of 2-5 K is
found in the 90 K bilayer of Fig. 6, but the corresponding dR/dT
peak is too small and broad for a clear determination of the
shift. Suppression in the magnitude of $dR/dT$ near the SRO peak
is evident in all bilayers in Figs. 6 and 7, and more so in the 90
K bilayer, similar to the results of Fig. 4.\\

Since the largest shifts of the ferromagnetic ordering temperature
$T_p$ are observed in the 30 and 60 K bilayers, it is tempting to
attribute this behavior to the fact that around 140K, YBCO with
$T_c$ of 30 K or 60 K is within its pseudogap regime, while the 90
K phase of YBCO is above its $T^*$. One possible interpretation is
that this phenomenon originates in a proximity effect where
preformed pairs with zero spin are injected from the YBCO layer
into the adjacent SRO layer and lower its $T_p$. To investigate
this scenario we performed several control experiments, in which
we compared the proximity effect between SRO and different types
of conductors, as follows: i. Bilayers of 5 nm SRO with 10 nm
thick cuprate films for which the relevant temperature range is
not in the pseudogap regime.  These include
$La_{1.88}Sr_{0.12}CuO_{4-y}$ (LSCO) \cite{Ofer} with $T^*<$100 K
and the 90 K YBCO phase for which the pseudogap regime, if at all,
exists only very close to $T_c$. ii. Bilayers of 5 nm SRO with 10
nm of a normal metal (Au) iii. Bilayers of 5 nm SRO with 10 nm of
an antiferromagnetic metal (Cr). In cases (i) and (ii), single
uncorrelated charge carriers are injected into the SRO, while in
case (iii) there is a competition between two different types of
magnetic order parameters. The proximity effect with an underdoped
YBCO should give similar results to case (iii), unless preformed
pairs exist. In the latter case, each injected pair carries two
charges with zero spin, which should decrease both the
polarization and the ferromagnetic correlations in the SRO layer
in a more noticeable way than in the case where the injected
charges are uncorrelated. One can argue qualitatively that if the
underdoped YBCO contains singlet Cooper pairs, the probability of
injecting two charges with zero spin into the SRO is
$P_{\uparrow\downarrow} = 1$ because of the electron-electron
correlations, while for uncorrelated electrons the injection
probability $P_{\uparrow\downarrow}$ would be only $P_\uparrow
P_\downarrow=0.5\times 0.5 = 0.25$. Thus clearly, the injection of
correlated pairs should have a stronger effect on $T_p$, the
midpoint of the ferromagnetic transition.\\

\begin{figure}
\includegraphics[height=6cm,width=8cm]{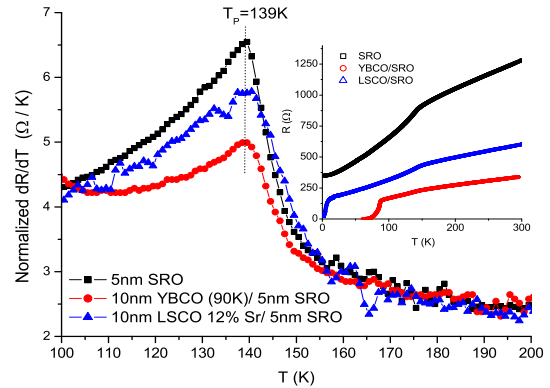}
\caption{\label{fig:epsart}(Color online) Normalized dR/dT versus
temperature of two bilayers, one of a 10 nm YBCO/5 nm SRO of the
90 K YBCO phase, and the other of a 10 nm
$La_{1.88}Sr_{0.12}CuO_{4-y}$/5 nm SRO, together with the dR/dT
data of a 5 nm thick SRO reference film. Normalization to the
reference film is done in the 160-180 K temperature range. The
inset shows the raw data of R versus T.}
\end{figure}

The dR/dT versus temperature results of the bilayers of the
control experiments (of which the LSCO data was obtained in the
configuration shown by the inset of Fig. 7), are shown in Figs. 8
and 9. This time the dR/dT peak of the 90 K YBCO bilayer in Fig. 8
is stronger and allows a clear determination of its peak value
$T_p$. Fig. 8 shows that the $T_p$ values of both the LSCO and 90
K YBCO bilayers coincide to within the experimental error with
that of the SRO reference film at $T_p\approx$139 K. Moreover, the
$T_p$ value of the Cr bilayer of Fig. 9 shifts by only 2 K to
lower temperature relative to the reference SRO film, while in the
Au bilayer a similar $T_p$ shift is observed but to higher
temperatures.  Thus, these $\pm$2 K $T_p$ shifts are significantly
smaller than the 6-8 K shifts observed in the 30 and 60 K bilayers
of Figs. 6 and 7.\\

\begin{figure}
\includegraphics[height=6cm,width=8cm]{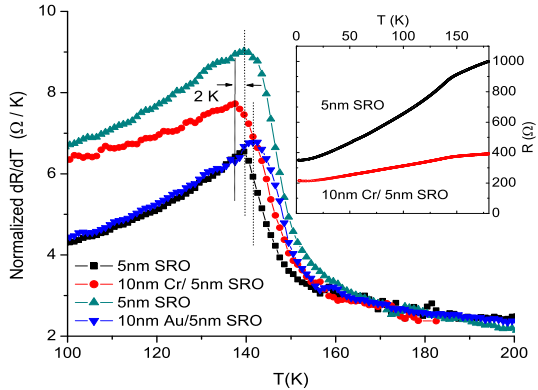}
\caption{\label{fig:epsart}(Color online) dR/dT versus temperature
of 10 nm Cr/5 nm SRO and 10 nm Au/5 nm SRO bilayers normalized to
two 5 nm thick SRO reference films. Normalization is done in the
160-180 K temperature range. The inset shows the raw data of R
versus T for the Cr bilayer and one of the SRO films.}
\end{figure}

We attribute the small 2 K shift down of $T_p$ of the Cr bilayer
in Fig. 9 to a loss of itinerant spin polarized electrons in the
SRO layer caused by an inverse proximity effect, where these
electrons are injected into the adjacent normal chromium layer.
Injection of electrons in the opposite direction into the SRO
layer should in principle enhance its ferromagnetism due to the
presence of the exchange field. This seems to be the case in the
Au/SRO bilayer in this Figure, where the gold has a higher density
of conduction electrons than SRO, but without spin correlations.
In the Cr/SRO bilayer, this is not the case since the carriers
injected into the SRO, although uncorrelated, may have some degree
of antiferromagnetic spin correlations which would tend to hinder
the formation of the ferromagnetic order ($T_p$) in the SRO. Fig.
9 shows exactly this for the chromium bilayer, by the small $T_p$
shift down of $\sim$2 K. Thus, injection of uncorrelated electrons
with opposite spins does not seem to cause the large shift down of
$T_p$. This leads to the conclusion that the large $T_p$ shifts
observed in Figs. 6 and 7 for the 30 and 60 K bilayers are due to
injection of correlated electrons, similarly to the preformed
pairs scenario in the pseudogap regime. One could still argue that
the large $T_p$ shifts down are due to the inverse proximity
effect, where a loss of itinerant spin polarized electrons occurs
as in the chromium case. This however, would necessitate that the
density of states near the Fermi surface of the 90 K YBCO phase be
lower than that of the 30 and 60 K phases of YBCO, which is not
the case. We note that the density of states in all the hole-doped
cuprates, and thus also in YBCO, increases with increasing oxygen
doping level. Thus one can rule out the inverse proximity effect
as the responsible mechanism for the large $T_p$ shifts. Our
results therefore, are consistent with the preformed pairs
scenario, but are not a definitive proof for their existence.
However, our suggestion that a large effect on $T_p$ is due to the
injection of oppositely polarized correlated electrons, from the
YBCO layer in the pseudogap regime into the SRO layer, is a
significant step in support of the precursor superconductivity
scenario in the cuprates.\\

Finally, we discuss the above conclusion and consider alternative
interpretations of our results. First, we wish discuss possible
interdiffusion effects at the interface on our results. From the
atomically smooth surface morphology of the SRO layer as seen in
Fig. 1, and the various TEM images
\cite{Char-TEM,Habermeier-Physica-C, Habermeier-PRB}, it is clear
that the interface in the bilayers with SRO is sharp and flat.
Therefore, interdiffusion in such an interface is quite unlikely.
Moreover, even if we assume that reaction products are present at
the interface, our results are not affected by them since in the
7.5 nm YBCO/5 nm SRO bilayers (see Fig. 4), no shift of $T_p$ was
observed. Thus the large $T_p$ shifts observed when the YBCO layer
was thicker (10 nm thick), have nothing to do with interface
reactions, if any, and result from the added amount of YBCO. We
also wish to stress that in both Cr/SRO and Au/SRO bilayers,
deposition of the Cr and Au layers was done at low temperatures
(30 and 150$^\circ$C, respectively). Hence, no interdiffusion into
the SRO could have occurred in these cases either. Concerning
density of states considerations, we note that the similar
resistivities of SRO, YBCO and LSCO indicates that the density of
states in these materials is on the same order of magnitude. The
density of states of the Cr and Au metals are much higher than
that of SRO, and as a result a higher injection rate into the SRO
is expected. This however, did not lead to larger $T_p$ shifts in
these cases. An alternative model for the pseudogap regime
involves large superconducting fluctuations above $T_c$, while
phase coherence is reached at $T_c$. Such a model is consistent
with the Nernst results where the onset of this effect follows a
higher temperature dome similar to the lower temperature dome at
$T_c$ \cite{Ong}. Our results for the 12\% Sr doped LSCO where no
$T_p$ shift was observed (Fig. 8), are consistent with the Nernst
onset at 110-120 K, since this is out of the ferromagnetic
transition range of SRO at 130-150 K. Unfortunately, there is no
similar data on the Nernst onset for YBCO, so no further
comparison with the YBCO results can be made. We note however that
our LSCO results are in disagreement with the pseudogap
$T^*\approx$160 K obtained from specific heat measurements
\cite{Ong}, but agree well with the $T^*\approx$60 K obtained from
STM results \cite{Ofer}.\\

In conclusion, using the ferromagnetic transition of SRO as a
probe, the YBCO properties in the pseudogap regime were studied in
nanometer thick bilayers with SRO. By comparing results of several
types of bilayers, we found that the largest effect on the SRO
layer occurs when the YBCO is in the pseudogap regime. We conclude
that in this regime in YBCO, electrons with opposite spins have to
be correlated to produce the much larger effect, and this lends
support for the preformed pairs scenario.\\

{\em Acknowledgments:}  The authors are grateful to L. Klein and
O. Millo  for useful discussions, and I. Asulin for taking the STM
image of Fig. 1. This research was supported in part by the Israel
Science Foundation (grants \# 1564/04 and 746/06), the Heinrich
Hertz Minerva Center for HTSC, the Karl Stoll Chair in advanced
materials, and by the Fund for the Promotion of Research at the
Technion.\\

\bibliography{AndDepBib.bib}

\bibliography{apssamp}

\end{document}